\documentstyle[eqsecnum,prd,aps,epsf]{revtex}

\draft
\begin{document}

\thispagestyle{empty}
{\baselineskip0pt
\leftline{\large\baselineskip16pt\sl\vbox to0pt{\hbox{\it Department of Physics}
               \hbox{\it Kyoto University}\vss}}
\rightline{\large\baselineskip16pt\rm\vbox to20pt{\hbox{KUNS 1476}
               \hbox{January, 1998} 
\vss}}%
}
\vskip1cm
\begin{center}{\large \bf
Neutron stars in scalar-tensor theories of gravity 
and catastrophe theory}
\end{center}
\vskip1cm
\begin{center}
 {\large 
Tomohiro Harada
\footnote{ Email address: harada@tap.scphys.kyoto-u.ac.jp}} \\
{\em Department of Physics,~Kyoto University,} 
{\em Kyoto 606-01,~Japan}\\
\end{center}

\begin{abstract}
We investigate neutron stars in scalar-tensor theories.
We examine their secular stability against 
spherically symmetric perturbations
by use of a turning point method.
For some choices of the 
coupling function contained in the theories,
the number of 
the stable equilibrium solutions changes
and the realized equilibrium solution may change 
discontinuously
as the asymptotic value of the scalar field
or total baryon number 
is changed continuously. 
The behavior of the stable equilibrium solutions
is explained by fold and cusp catastrophes.
Whether or not the cusp catastrophe appears
depends on the choice of
the coupling function.
These types of catastrophes are structurally stable.
Recently discovered spontaneous scalarization,
which is a nonperturbative strong-field phenomenon due
to the presence of the
gravitational scalar field,
is well described in terms of the cusp
catastrophe.
\end{abstract}

\section{Introduction}
Scalar-tensor theories~\cite{will1993,de1992} are among 
the generalized theories of gravitation.
Brans-Dicke theory~\cite{bd1961} 
is a member of the
scalar-tensor theories.
Scalar-tensor theories have recently
attracted the attention of many researchers.
One of the reasons is that the unified theories
that contain gravity as well as other interactions,
such as string theory~\cite{gsw1987}, 
naturally predict the existence of scalar fields
that relate to gravity.
In the hyperextended inflation 
model~\cite{sa1990}, 
scalar-tensor theories of gravity
play an essential role.
Moreover, projects of laser interferometric gravitational wave 
observations~\cite{ligo,virgo,geo,tama} 
will be soon in practical use, so that high-accuracy
tests of the scalar-tensor theories may be 
expected~\cite{wz1989,will1994,snn1994,sst1995,hcnn1997}.

Scalar-tensor theories are viable theories of gravity
for some choices of the coupling function which
is contained in the theories.
Predictions of these theories in a strong field may be drastically
different from those of general relativity.
Recently, Damour and 
Esposito-Far\`ese~\cite{de1993,de1996} discovered
one example of such phenomena.
They showed that, for some choices of the
coupling function,
the configuration of a massive neutron star 
deviates significantly from that in general relativity,
even if the post-Newtonian limit of the theory 
is extremely close to
or even agrees with that of general relativity.
This deviation in a strong field 
may be easily tested from binary-pulsar timing
observations, if it exists,
because of the extra energy loss 
by scalar gravitational radiation~\cite{de1996}.
The deviation from general relativity 
can be no longer dealt with as a perturbative effect
from general relativity.
Damour and Esposito-Far\`ese referred to this nonperturbative
strong-field effect as ``spontaneous scalarization''
in analogy to the spontaneous magnetization
of the ferromagnets.

In this paper, we investigate spontaneous scalarization
in detail with the technique of catastrophe theory.
A many-parameter version of the turning point 
method~\cite{katz1978,katz1979,sorkin1981,sorkin1982}
is used
as a tool of a stability analysis of equilibrium solutions.
The stability analysis of boson stars in scalar-tensor
gravity via catastrophe theory
in the case of one-dimensional control space
was done in~\cite{cs1997,tls1997}.
Catastrophe types of 
neutron star equilibrium solutions 
are classified as fold 
and cusp catastrophes.
The occurrence of the cusp catastrophe 
depends on the choice of the coupling function.
Spontaneous scalarization is classified as
the cusp catastrophe.
From this catastrophic feature, we conclude that 
the stable configuration of the neutron star
may change discontinuously as the baryon number
of the star or the asymptotic value of 
the scalar field changes continuously.
The behavior of the scalar charge
around the cusp point 
is explained by catastrophe theory.
For the coupling function considered here, 
when the asymptotic value of the scalar field
is such that the theory agrees with general relativity
in the post-Newtonian limit, we find
the sequence of the equilibrium solutions
{\it bifurcates} to 
three branches at some critical central density.
One branch consists of solutions
that are identical to neutron stars in general relativity,
and the other two consist
of solutions that deviate significantly
from neutron stars in general relativity. 
The general relativistic branch is secularly unstable
in agreement with the result obtained by 
a perturbation study~\cite{harada1997},
while the non-general-relativistic branches are secularly stable.

This paper is organized as follows.
In Sec. II, we summarize the field equations of 
scalar-tensor theory
and the equations determining equilibrium solutions
of neutron stars in this gravitational theory.
In Sec. III, we present stability criteria
on the grounds of the turning point method.
In Sec. IV, the stability criteria are
applied to equilibrium solutions of neutron stars in 
scalar-tensor theory
and some consequences of catastrophe theory
are discussed.
Section V is devoted to conclusions.
We use units in which $c=1$.
The Greek indices run from 0 to 3.
We follow the Misner-Thorne-Wheeler~\cite{mtw1973} sign conventions
for curvature quantities.
  
\section{Basic equations}
Here we consider a class of scalar-tensor theories
in which gravity is mediated by not only a metric tensor
but also a massless scalar field.
The action is given by~\cite{de1992}
\begin{equation}
  \label{eq:action}
    I=\frac{1}{16\pi G_*}\int\sqrt{-g_*}\left( R_* -
    2g_*^{\mu\nu}\varphi_{,\mu}\varphi_{,\nu}
    \right)d^4x
    +I_m[\Psi_m,A^2(\varphi)g_{*\mu\nu}], 
\end{equation}
where $g_{*\mu\nu}$ is the ``Einstein'' frame
metric tensor, 
$\Psi_m$ denotes matter fields collectively, and $G_*$ is some
dimensionful constant.
In this Einstein frame, the Einstein-Hilbert term
is isolated from other sectors.
The ``Brans-Dicke'' frame metric tensor $\tilde{g}_{\mu\nu}$ 
is related to the Einstein frame metric tensor by
the following conformal transformation: 
\begin{equation}
      \tilde{g}_{\mu\nu}=A^2(\varphi)g_{*\mu\nu}.
      \label{eq:conformaltransformation}
\end{equation}
Because of the ``universal coupling,'' which is the way of the 
coupling of the scalar field in the matter sector seen
in Eq. (\ref{eq:action}),
a test particle moves on the geodesic of the Brans-Dicke 
frame metric $\tilde{g}_{\mu\nu}$.
For this reason the Brans-Dicke frame is often called
a ``physical'' frame.
The tilde denotes the physical frame quantity.

In the Einstein frame, the field equations are given by
\begin{eqnarray}
      & & G_{*\mu\nu}=8\pi G_* T_{*\mu\nu}
      +2\left(\varphi_{,\mu}\varphi_{,\nu}-\frac{1}{2}g_{*\mu\nu}
      g_*^{\alpha\beta}\varphi_{,\alpha}\varphi_{,\beta}\right), 
    \label{eq:fieldeq1} \\
      & & \Box_* \varphi = - 4\pi G_* \alpha(\varphi) T_*,
      \label{eq:fieldeq2}
\end{eqnarray}
while the equations of motion for matter are 
\begin{equation}
  \label{eq:eom}
  \nabla_{*\nu}T_{*\mu}^{\nu}=\alpha(\varphi)T_*\nabla_{*\mu}\varphi,
\end{equation}
where the energy-momentum tensor of the matter, $T_*^{\mu\nu}$,
is defined and related to the physical energy-momentum
tensor $\tilde{T}^{\mu\nu}$ as
\begin{equation}
  \label{eq:defofem}
      T_*^{\mu\nu} \equiv \frac{2}{\sqrt{-g_*}}
      \frac{\delta I_m[\Psi_m,A^2(\varphi)g_{*\mu\nu}]}
      {\delta g_{*\mu\nu}}
      = A^6(\varphi) \tilde{T}^{\mu\nu}.
\end{equation}
$G_{*\mu\nu}$ and $\Box_*$ are the Einstein tensor and
d'Alembertian of $g_{*\mu\nu}$, respectively.
$T_*$ and $\alpha(\varphi)$ are defined as
\begin{eqnarray}
      T_* &\equiv& T_{*\mu}^{~~\mu} 
      \equiv T^{\mu\nu}_* g_{*\mu\nu},\\
      \alpha(\varphi) &\equiv& \frac{d\ln A(\varphi)}
      {d\varphi}.
\end{eqnarray}

The parameters in the parametrized post-Newtonian framework are
given by~\cite{will1993,de1992}
\begin{eqnarray}
      & & 1-\gamma_{Edd} = \frac{2\alpha_0^2}{1+\alpha_0^2}, \\
      & & \beta_{Edd}-1 = \frac{\beta_0\alpha_0^2}{2(1+\alpha_0^2)^2},\\
      & & \xi = \alpha_1 = \alpha_2 = \alpha_3 = 0,   
\end{eqnarray}
where $\beta_{Edd}$ and $\gamma_{Edd}$ are
the so-called Eddington parameters.
We have defined
\begin{eqnarray}
  \label{eq:defofalpha0}
      \alpha_0 &\equiv& \alpha(\varphi_0), \\
      \label{eq:defofbeta0}
      \beta_0 &\equiv& \frac{d\alpha}
      {d\varphi}(\varphi_0),
\end{eqnarray}
and
$\varphi_0$ is the value of the scalar field
$\varphi$ in the 
spatial asymptotic region.
We assume that the cosmological evolution of the scalar field
is sufficiently slow in comparison with the characteristic
time scale of the local gravitational process of the
isolated object considered here. 
From this assumption 
$\varphi_0$ is regarded as the cosmological value of the scalar field.
On the other hand, we can identify the asymptotic value $\varphi_0$
to the value of the scalar field in the matching region
in the matching approach to the $N$-compact-body problem
(see Appendix A of~\cite{de1992}). 
Then, the solar-system experimental constraints are~\cite{lebach1995}
\begin{equation}
  \label{eq:constraint1}
      \gamma_{Edd}= 0.9996\pm0.0017
\end{equation}
and~\cite{williams1995}
\begin{equation}
  \label{eq:constraint2}
  4\beta_{Edd}-\gamma_{Edd}-3=-0.0007\pm0.0010.
\end{equation}

We summarize equations for the structure of a relativistic star
in scalar-tensor theory, following~\cite{de1993}.
We restrict ourselves to 
the static and spherically symmetric case.
The metric is given in the following form:
\begin{equation}
  \label{eq:staticspherical}
  ds_{*}^2 = - e^{\nu(r)}dt^2 + \left(1-\frac{2\mu(r)}{r}
    \right)^{-1}dr^2
  +r^2(d\theta^2+\sin^2 \theta d\phi^2).
\end{equation}
The matter 
is described as a perfect fluid: i.e.,
\begin{equation}
  \tilde{T}_{\mu\nu}=(\tilde{\rho}+\tilde{p})\tilde{u}_{\mu}
  \tilde{u}_{\nu}+\tilde{p}\tilde{g}_{\mu\nu}.
\end{equation}
Then,
the following equations are obtained:
\begin{eqnarray}
  \label{eq:beq1}
  \mu^{\prime}&=&4\pi G_* r^2 A^4 \tilde{\rho}
  +\frac{1}{2}r(r-2\mu)\psi^2, \\
  \label{eq:beq2}
  \nu^{\prime}&=&8\pi G_* \frac{r^2 A^4 \tilde{p}}{r-2\mu}
  +r\psi^2+\frac{2\mu}{r(r-2\mu)}, \\
  \label{eq:beq3}
  \varphi^{\prime}&=&\psi, \\
  \label{eq:beq4}
  \psi^{\prime}&=&4\pi G_* \frac{r A^4}{r-2\mu}
  [\alpha(\tilde{\rho}-3\tilde{p})+r(\tilde{\rho}-\tilde{p})\psi]
  -\frac{2(r-\mu)}{r(r-2\mu)}\psi, \\
  \label{eq:beq5}
  \tilde{p}^{\prime}&=&-(\tilde{\rho}+\tilde{p})
  \left[4\pi G_*\frac{r^2 A^4 \tilde{p}}{r-2\mu}
    +\frac{1}{2}r\psi^2+\frac{\mu}{r(r-2\mu)}
  +\alpha(\varphi)\psi\right], \\
  \label{eq:beq6}
  \tilde{p}&=&\tilde{p}(\tilde{\rho}),
\end{eqnarray}
where the prime denotes a derivative with respect to $r$.
We use the polytropic 
equations of state:
\begin{eqnarray}
  \tilde{\rho} &=& \tilde{n}m_b+\frac{Kn_0m_b}{\Gamma-1}
  \left(\frac{\tilde{n}}{n_0}\right)^{\Gamma}, \\
  \tilde{p} &=& Kn_0m_b\left(\frac{\tilde{n}}{n_0}\right)^{\Gamma}, \\
  m_b &=& 1.66\times10^{-24}~\mbox{g}, \\
  n_0 &=& 0.1~\mbox{fm}^{-3},
\end{eqnarray}
where $\tilde{n}$ is the baryon number density in the 
Brans-Dicke frame.
We then take the parameters $\Gamma=2.34$ and
$K=0.0195$
(EOS II of~\cite{de1993}).
Note that the total baryon number is
given by
\begin{equation}
  N=\int 4\pi\tilde{n} A^3 r^2 \left(1-\frac{2\mu}{r}
  \right)^{-1/2}dr.
\end{equation}

Here we present the method of solving the above equations
and obtaining the structure of a neutron star.
First the initial values of the above set of ordinary differential
equations are fixed as
\begin{equation}
  \mu (0)=0,~\nu(0)=0,~\varphi(0)=\varphi_c,
  ~\psi(0)=0,~\tilde{p}(0)=\tilde{p}_c,
\end{equation}
and Eqs. (\ref{eq:beq1})-(\ref{eq:beq6}) are integrated
numerically up to the stellar surface at which $\tilde{p}=0$.
Thereafter the solution is matched with the static and
spherically symmetric ``vacuum'' solution, where 
the term ``vacuum'' 
means only the absence of matter, i.e., $\tilde{T}_{\mu\nu}=0$.
This solution is given in~\cite{de1992}.
The solutions are parametrized by three parameters
$b$, $d$, and $\varphi_0$.
From the matching conditions at the surface, we can obtain 
$\nu(r)$ including the constant term.
In order to set $A(\varphi_0)$ to unity,
we rescale the raw quantities
to the renormalized ones
as follows:
\begin{eqnarray}
  r_{ren}&=&A_0^2 r,~~ \mu_{ren}= A_0^2 \mu,~~ \nu_{ren}=\nu,
  ~~\varphi_{ren}=\varphi,~~ \psi_{ren}=A_0^{-2}\psi, \nonumber \\
  \varphi_{0 ren} &=& \varphi_0,~~ a_{ren}=A_0^2 a,
  ~~b_{ren}=A_0^2 b, ~~d_{ren}=A_0^2 d, ~~N_{ren}=A_0^3 N.
\end{eqnarray}
Then,
from the asymptotic properties at spatial infinity for a static 
and isolated system
\begin{eqnarray}
  g_{*\mu\nu}&=&\eta_{\mu\nu}+\frac{2G_*m}{r_{ren}}\delta_{\mu\nu}
  +O\left(\frac{G_*^2}{r_{ren}^2}\right), \\
  \varphi&=&\varphi_0+\frac{G_*\omega}{r_{ren}}
  +O\left(\frac{G_*^2}{r_{ren}^2}
    \right), 
\end{eqnarray}
where $\eta_{\mu\nu}$ is the Minkowskian metric.
We call $m$ the Arnowitt-Deser-Misner (ADM) 
energy and $\omega$ the scalar charge~\cite{de1992}.
$b_{ren}$ and $d_{ren}$ are related to $m$ and $\omega$ as
\begin{eqnarray}
  G_* m=\frac{b_{ren}}{2}, \\
  G_*\omega= -d_{ren}.
\end{eqnarray}
Hereafter the subscripts ``{\it ren}'' are omitted for simplicity. 
We use units in which $G_*=1$.
  
\section{stability criteria}

In scalar-tensor theory, control parameters of the static, 
spherically symmetric, and isolated neutron star are not only 
the baryon number $N$
but also the ``external field'', that is, 
the asymptotic value of 
the scalar field $\varphi_0$. 
For static systems, the partial derivative of $m$ 
in terms of $\varphi_0$  
with $N$ constant is given by~\cite{de1992}
\begin{equation}
  \label{eq:dermderphi0}
  \left(\frac{\partial m}{\partial \varphi_0}\right)_{N}=-\omega.
\end{equation}
The energy injection of the system by 
increasing baryons is described as
\begin{equation}
\label{eq:injection}
  \int 4\pi \tilde{u}_0 \delta\tilde{\rho}
  A^3 r^2 \left(1-\frac{2\mu}{r}\right)^{-1/2}dr 
  =\int 4\pi e^{\nu/2} \tilde{\mu}
  \delta \tilde{n}
  A^4 r^2 \left(1-\frac{2\mu}{r}\right)^{-1/2}dr,
\end{equation}
where $\tilde{\mu}\equiv d\tilde{\rho}/d\tilde{n}$
is the chemical potential.
The first law of thermodynamics
in an adiabatic process is
\begin{equation}
  \label{eq:1stlaw}
  d\left(\frac{\tilde{\rho}}{\tilde{n}}\right)
    =-\tilde{p}d\left(\frac{1}{\tilde{n}}\right).
\end{equation}
From Eqs. (\ref{eq:beq2}), (\ref{eq:beq3}),
(\ref{eq:beq5}), and (\ref{eq:1stlaw}),
we find that the quantity $A e^{\nu/2}\tilde{\mu} $ 
is constant all over the star.
Therefore this quantity can be estimated  
by its value at the stellar surface.
Using this fact, the expression of the energy injection, 
Eq. (\ref{eq:injection}), is rewritten as
\begin{equation}
  A_s e^{\nu_s/2}\tilde{\mu}_s\int 4\pi A^3 r^2
  \left(1-\frac{2\mu}{r}\right)^{-1/2}\delta \tilde{n}dr 
  = A_s e^{\nu_s/2}\tilde{\mu}_s \delta N,
\end{equation}
where the suffix ``{\it s}'' indicates that the quantity is evaluated 
at the stellar surface $r=r_s$.
Therefore, 
the effective chemical potential, $\mu_{eff}$, is given by
\begin{equation}
  \mu_{eff}\equiv \left( \frac{\partial m}{\partial N} 
    \right)_{\varphi_0}
  = A_s e^{\nu_{s}/2}\tilde{\mu}_{s}.
\end{equation}
From the above discussions the variation of $m$ for
static systems in a quasistatic process is written
in the following form:
\begin{equation}
  \delta m = -\omega \delta\varphi_0 + \mu_{eff} \delta N, 
\end{equation}
where by ``quasistatic process'' we mean
successive changes among the
infinitesimally nearby equilibrium solutions.

Suppose that the isolated neutron star is 
perturbed slightly with $\varphi_0$ and $N$ constant
for some reason other than incident
waves, while spherical symmetry is
preserved.
Then the outgoing waves can carry out
some positive energy to infinity and
the system cannot keep its original state
if there exists an energetically favorable
configuration which is infinitesimally 
deformed from the system with the same
$\varphi_0$ and $N$.
Therefore, an equilibrium solution $X$
is secularly stable against 
spherically symmetric (infinitesimal)
perturbations if and only if there is no 
momentarily static and
spherically symmetric configuration $Y$ which is
arbitrarily close to $X$ with the same $\varphi_0$ and $N$
but strictly smaller $m$.

In order to examine the stability of the 
equilibrium solution, we follow the turning point
method~\cite{katz1978,katz1979,sorkin1981,sorkin1982}.
In the present problem, $m$ is a potential
function,
since the equilibrium solution is a stationary point
of $m$ ($\delta m=0$) and 
the stable equilibrium solution is a minimal point of $m$
($\delta m=0$ and $\delta^2 m >0$).
The asymptotic value of the scalar field $\varphi_0$
and baryon number $N$ form a two-dimensional control space.
The equilibrium solutions are uniquely parametrized
by two parameters, i.e., 
the central value of the scalar field, $\varphi_c$, and
the central total baryonic density, $\tilde{\rho}_c$.

We adopt the following stability criteria~\cite{sorkin1982}:

(i) The stability of $X(\varphi_c,\tilde{\rho}_c)$ 
can change typically 
only at a ``turning point.''
Here the ``turning point'' $(\varphi_c^0,\tilde{\rho}_c^0)$ 
is a point where
there exists a nontrivial vector 
$(\delta\varphi_c,\delta\tilde{\rho}_c)$ 
such that 
\begin{eqnarray}
  \label{eq:tp1}
  \delta \varphi_0 &=& \left(\frac{\partial \varphi_0}{\partial
  \varphi_c}\right)_{\tilde{\rho}_c}
  \delta \varphi_c + \left(\frac{\partial \varphi_0}{\partial
  \tilde{\rho_c}}\right)_{\varphi_c}
  \delta \tilde{\rho}_c=0,\\
  \label{eq:tp2}
  \delta N &=& \left(\frac{\partial N}{\partial \varphi_c}
  \right)_{\tilde{\rho}_c}  
  \delta \varphi_c +  \left(\frac{\partial N}{\partial \tilde{\rho_c}}
  \right)_{\varphi_c}
  \delta \tilde{\rho}_c=0.
\end{eqnarray}
From Eqs. (\ref{eq:tp1}) and (\ref{eq:tp2}), 
\begin{equation}
\frac{\partial(\varphi_0,N)}{\partial(\varphi_c,\tilde{\rho}_c)}=0
\end{equation}
at the turning point.
Therefore the change of stability can be detected as 
envelopes of a family of curves $\tilde{\rho}_c=\mbox{const}$
in the $(\varphi_0, N)$ plane.
Of course, this is also true for a family of
curves $\varphi_c=\mbox{const}$.

(ii) In order to specify an unstable branch at the
turning point,
we draw the sequence of equilibrium solutions 
in the $(\varphi_0,\omega)$ plane,
maintaining $N$ constant.
Then, as one proceeds along the curve 
in a counterclockwise direction,
a branch beyond the turning point is unstable.
This is also the case with the curve $(N,-\mu_{eff})$
with $\varphi_0$ constant.
This is a direct consequence of theorem I of
~\cite{sorkin1982}.

Here we describe the meaning of criterion (i) 
in the context of catastrophe theory.
We regard the ADM energy $m$ as a function
of three variables $\varphi_0$, $N$, and $\omega$.
We take $\omega$ as a state variable. 
We consider an equilibrium space
\begin{equation}
  M_m=\left\{(\varphi_0,N,\omega)
      \Biggm|\left(\frac{\partial m}{\partial \omega}
    \right)_{\varphi_0,N}=0\right\}
\end{equation}
and a control space
\begin{equation}
  {\bf R}^2=\{(\varphi_0,N)\}.
\end{equation}
We define a catastrophe map
\begin{eqnarray}
  \chi_m : M_m~ & & \longrightarrow~ {\bf R}^2, \nonumber \\ 
           (\varphi_0, N, \omega)~ & & \longmapsto ~(\varphi_0,N).  
\end{eqnarray}
A point $P \in M_m$ is called a singular point of $\chi_m$
if the Jacobian of $\chi_m$ vanishes at $P$.
A point $Q \in {\bf R}^2$ is called a singular value 
if there is at least one singular point in $\chi_m^{-1}(Q)$.
A bifurcation set $B_m\subset {\bf R}^2$ is 
a set of singular values.
At a singular point $P\in M_m$, a
vector normal the tangent space of $M_m$, which is
\begin{equation}
  \left(\left(\frac{\partial^2 m}
      {\partial\varphi_0\partial\omega}\right)
  _{N,\omega},
  \left(\frac{\partial^2 m}{\partial N\partial\omega}\right)
  _{\omega,\varphi_0},
  \left(\frac{\partial^2 m}{\partial\omega^2}\right)
  _{\varphi_0,N}\right),
\end{equation}
is parallel to the $\varphi_0 N$ plane.
Therefore, the set of singular points,
$\Sigma_m\subset M_m$,
satisfies
\begin{equation}
  \Sigma_m=\left\{(\varphi_0,N,\omega)\Biggm|
      \left( \frac{\partial m}{\partial \omega}
        \right)_{\varphi_0,N}=\left(\frac{\partial^2 m}
        {\partial \omega^2}
        \right)_{\varphi_0,N}=0\right\},
\end{equation}
and the bifurcation set $B_m \subset {\bf R}^2$
satisfies
\begin{equation}
  B_m = \left\{(\varphi_0,N)\Biggm|
    \left( \frac{\partial m}{\partial \omega}
        \right)_{\varphi_0,N}=\left(\frac{\partial^2 m}
        {\partial \omega^2}
        \right)_{\varphi_0,N}=0\right\}.
\end{equation}
The envelopes of the family of the
curves $\tilde{\rho}_c=\mbox{const} $ 
in the $(\varphi_0,N)$ plane form a bifurcation set $B_m$  of the 
catastrophe map $\chi_m$ because the Jacobian of 
$\chi_m $ vanishes at points on the envelopes.
Criterion (i) says that 
a sequence of the equilibrium solutions
can change its stability only at the points
of the bifurcation set.

From criteria (i) and (ii), we examine the stability of the equilibrium solutions
of neutron stars in scalar-tensor theory. 
From the turning point method alone, however, we
cannot say that an equilibrium solution is {\it stable}.
Therefore the stability of an equilibrium solution must be investigated 
by perturbation study {\it once for all}.
For this purpose, we examine the case in which $\alpha(\varphi_0)=0$
and $m/r_s$ is sufficiently small ($\tilde{\rho}_c$ is
sufficiently small).
In this case, there is an equilibrium solution that is identical to that 
in general relativity.
For this solution,
the second-order variation of $m$ by
regular, adiabatic, 
time-symmetric, and spherically symmetric perturbations
with $\varphi_0$ and $N$ constant
is 
\begin{equation}
  \delta^2 m = \mbox{general relativistic terms} +
  \frac{1}{2}
  \int_{0}^{\infty} dr e^{-\nu/2}\left(1-\frac{2\mu}{r}\right)^{-1/2}
  \zeta \left[-\frac{d^2}{dr_*^2}+V(r)\right]\zeta,
\end{equation}
where 
\begin{eqnarray}
  \zeta&\equiv& r\delta\varphi, \\
  dr_* &\equiv& e^{\nu/2}\left(1-\frac{2\mu}{r}\right)^{1/2}dr, \\
  V(r) &\equiv& \frac{1}{r}\left(1-\frac{2\mu}{r}\right)\left[
    \frac{\nu^{\prime}}{2}-\frac{\mu^{\prime}r-\mu}{r(r-2\mu)}\right]
  e^{\nu}-4\pi \beta_0 (-\tilde{\rho}+3\tilde{p})e^{\nu},
\end{eqnarray}
and see Appendix B of~\cite{htww1964}
for the general relativistic terms.
The general relativistic part is positive definite
if $\Gamma>\Gamma_c$, and $\Gamma_c \to 4/3$ in the
Newtonian limit~\cite{chandrasekhar1964}. 
The second term is positive definite
because the eigenvalues of the operator, $-d^2/dr_*^2+V$,
are all positive 
for an arbitrary 
coupling function $A(\varphi)$
if $m/r_s$ is sufficiently small,
which has been shown in~\cite{harada1997}.
Therefore, the general relativistic equilibrium solution
in which the central density is sufficiently small
is stable for the case $\alpha(\varphi_0)=0$
if $\Gamma > \Gamma_c \simeq 4/3$.

\section{Results}

Hereafter we restrict our attention
to the coupling function of the quadratic form
\begin{equation}
  \label{eq:quadratic}
  A(\varphi)=\exp\left(\frac{1}{2}\beta\varphi^2\right).
\end{equation}
For this model, the solar-system experiments constrain
the present cosmological value of the scalar field 
through Eqs. (\ref{eq:constraint1}) and 
(\ref{eq:constraint2}) as
\begin{equation}
  |\varphi_0|\alt 0.032|\beta|^{-1} 
\end{equation}
and
\begin{equation}
  |\varphi_0|\alt\cases{
    0.012(1+\beta)^{-1/2}|\beta|^{-1} & for $\beta>-1 $, \cr
    0.029|1+\beta|^{-1/2}|\beta|^{-1} & for $\beta<-1 $, \cr
    }
\end{equation}
respectively.
In particular, if $\varphi_0=0$,
the post-Newtonian limit of this theory
agrees completely with that of general relativity
because $\alpha(\varphi_0)=0$.

\subsection{$\beta\protect\agt-4.35$ case}
We present here the results of the case $\beta=-4$,
but the features are basically common to
the case $\beta\agt-4.35$.
Figure 1 shows $\tilde{\rho}_c=\mbox{const}$ curves
in the $(\varphi_0,N)$ plane,
where the equilibrium solutions have been determined in the
manner described in Sec. II.
At a point on an envelope of the family of the curves 
seen in Fig. 1, the stability of the
sequence of equilibrium solutions changes.
Figure 2 shows the curves $(\varphi_0,\omega)$
with $N$ constant.
In Fig. 2, the solid lines denote stable branches
while the dotted lines denote unstable branches,
where stability criteria (i) and (ii) are applied.
Therefore, in region (A) in Fig. 1, 
only one stable equilibrium solution exists.
For $\varphi_0=0$,
this stable solution is identical to that of general relativity.
In region (B) in Fig. 1, however, no stable solution exists.
This is classified as the fold catastrophe in which
the control space is two-dimensional.
This catastrophe is elementary and structurally stable.
Hence it is expected that 
this catastrophe structure is not changed by adding 
small higher-order terms to the exponent of the coupling 
function (\ref{eq:quadratic}).
The potential function $m$ is written
locally around point $p(\varphi_{0p},N_{p})$ 
on the envelope (see Fig. 1) as, for $\varphi_{0p}>0$,
\begin{equation}
  \label{eq:fold}
  m = \frac{A}{3}(\omega-\omega_{p})^3
  +[B(\varphi_{0p}-\varphi_0)+B^{\prime}(N-N_{p})]
  (\omega-\omega_{p})+m_p, 
\end{equation}
where $A$, $B$, and $B^{\prime}$ are some positive constants.
Then the terms in the square brackets cancel out
on the envelope, and are negative in region (A)
and positive in region (B).
For $\varphi_{0p}<0$, 
replace $\omega-\omega_p$ and $\varphi_{0p}-\varphi_0$ with
$\omega_{p}-\omega$ and $\varphi_{0}-\varphi_{0p}$, respectively.
For simplicity, we describe the behavior of the scalar charge 
for the case $\varphi_{0p}>0$. 
From Eq. (\ref{eq:fold}), near 
point $p$, the scalar charge is given by
the roots of the following quadratic equation:
\begin{equation}
  \left(\frac{\partial m}{\partial \omega}\right)_{\varphi_0,N}
  =A (\omega-\omega_{p})^2 +
  [B(\varphi_{0p}-\varphi_0)+B^{\prime}(N-N_{p})]=0.
\end{equation}
The scalar charge is then given 
near point $p$ in region (A) by
\begin{equation}
  \omega = \omega_{p}\pm A^{-1/2}
  [B(\varphi_{0}-\varphi_{0p})+B^{\prime}(N_{p}-N)]^{1/2},
\end{equation}
where the upper sign denotes the stable branch and the lower denotes
the unstable one.
If a quadratic term in $(\omega-\omega_p)$ was involved in
Eq. (\ref{eq:fold}),
the number of the roots of the equation $\partial m/\partial\omega=0$
did not change at point $p$.
That is why Eq. (\ref{eq:fold}) does not contain the quadratic
term.
The ``scalar susceptibility'' $\chi_{\varphi}$ is given 
near point $p$ by
\begin{equation}
  \chi_{\varphi}\equiv\left(\frac{\partial \omega}{\partial \varphi_0}
    \right)_{N}
  =\pm\frac{1}{2}A^{-1/2}B
  [B(\varphi_{0}-\varphi_{0p})+B^{\prime}(N_{p}-N)]^{-1/2}.
\end{equation}
The bifurcation set $B_m\subset {\bf R}^2$, 
which is the envelope seen in Fig. 1,
is given by
\begin{eqnarray}
  \left(\frac{\partial m}{\partial \omega}\right)_{\varphi_0,N}
  &=&A (\omega-\omega_{p})^2 +
  [B(\varphi_{0p}-\varphi_0)+B^{\prime}(N-N_{p})]=0, \\
  \left(\frac{\partial^2 m}{\partial \omega^2}\right)
  _{\varphi_0,N}
  &=& 2A(\omega-\omega_p)=0,
\end{eqnarray}
i.e.,
\begin{equation}
  \varphi_0=\frac{B^\prime}{B}(N-N_p)+\varphi_{0p},
\end{equation}
near point $p$.
This fold catastrophe appears also
in general relativity in which $A(\varphi)=1$ identically.
In general relativity, because of the
absence of a gravitational scalar field, 
the control space is one-dimensional. 
For $\beta=-4$, the maximum ADM energy is 
greater than the general relativistic
one for $\varphi_0\neq 0$.
This is because, due to the presence of the scalar field,
the effective gravitational constant becomes smaller and
thereby gravity becomes weaker than in general relativity. 

\subsection{$\beta\protect\alt -4.35$ case}

This case is more interesting than the above case.
We present the results of the case $\beta=-6$.
Figure 3 shows $\tilde{\rho}_c=\mbox{const}$ curves
in the $(\varphi_0,N)$ plane.
This figure is very different from Fig. 1.
On the envelope of the family of curves,
$e^{\prime}dcbab^{\prime}c^{\prime}de$,
the sequence of equilibrium solutions changes its stability.
Although there are other envelopes in region (B),
they have nothing to do with the change of the number of stable
equilibrium solutions.
Figure 4 shows curves $(\varphi_0,\omega)$
with $N$ constant.
From criteria (i) and (ii),
the number of stable equilibrium solutions is as follows:
In region (A), only one stable equilibrium solution exists.
In region (B), two distinct stable equilibrium solutions exist.
Surprisingly,
these stable equilibrium solutions are different 
even for $\varphi_0=0$
from their counterparts in general relativity.
For $\varphi_0=0$, the unstable solution 
agrees with the stable solution
in general relativity.
One of the two stable equilibrium solutions disappears 
on the envelope $dcbab^{\prime}c^{\prime}d$.
In region (C), no stable equilibrium solution exists.
Point $a$ is a bifurcation point for $\varphi_0=0$.
This is seen in Fig. 5 which displays
the curves $(\tilde{\rho_c},m)$ 
and $(\tilde{\rho_c},N)$ for $\varphi_0=0$,
where the solid lines denote stable branches
and the dotted lines denote unstable branches.
In this figure, two stable branches are degenerate because,
for $\varphi_0=0$, two stable equilibrium solutions
are identical except for the sign of the scalar field.
The equilibrium solution of the bifurcated stable 
branches is more compact for 
smaller mass but less compact for larger mass
than the general relativistic sequence.
Figure 6 shows the equilibrium space $M_m$
near point $a$.
This type of the catastrophe at point $a$ is classified 
as the cusp catastrophe in which
the control space is two-dimensional.
The map $\chi_m$ is a cusp catastrophe map.
This catastrophe is elementary and structurally stable, 
which suggests that
this structure is stable against adding small
higher order terms to the exponent of the coupling 
function (\ref{eq:quadratic}).
Point $a$ is called a cusp point.

We restrict our attention to cusp point $a(0,N_a)$.
($m_b N_a \simeq 1.24 M_{\odot}$ for $\beta=-6$.)
The potential function $m$
is written around cusp point $a$ as
\begin{equation}
  \label{eq:cusp}
  m=\frac{C}{4}\omega^4-\frac{D(N-N_a)}{2}\omega^2
  -\varphi_0\omega+m_a,
\end{equation}
where $C$ and $D$ are some positive constants.
The reason why the coefficient of $\varphi_0 \omega$ is
determined is that Eq. (\ref{eq:dermderphi0}) holds.
This form of expansion agrees with the usual Landau ansatz
for a second-order phase transition, which
has been used to explain spontaneous scalarization
by Damour and Esposito-Far\`ese~\cite{de1996}.
The scalar charge is given by the roots of
the following cubic equation:
\begin{equation}
  \label{eq:cube}
  \left(\frac{\partial m}{\partial \omega}\right)
  _{\varphi_0,N}=
  C\omega^3-D(N-N_a)\omega-\varphi_0=0.
\end{equation}
From Eq. (\ref{eq:cube}), near cusp point $a$, 
the scalar charge is given by
\begin{equation}
  \omega=0,
\end{equation}
for $N<N_a$ with $\varphi_0=0$. 
This is a stable branch. For $N>N_a$ with $\varphi_0=0$,
\begin{equation}
  \label{eq:critical}
  \omega=\cases{
    \pm\left(\frac{D}{C}\right)^{1/2}(N-N_a)^{1/2} 
    & for the stable branches, \cr
    0 & for the unstable branch. \cr }
\end{equation}
At point $a$ the stable equilibrium solution changes
{\it continuously}, but its derivative
with respect to $N$ is {\it discontinuous}.
If Eq. (\ref{eq:cusp}) involved a cubic term in $\omega$,
the number of roots of the equation $\partial m/\partial \omega=0$
did change at point $a$.
But this catastrophe was classified as the fold type
and therefore not the case for point $a$
because of the shape of the bifurcation set seen in Fig. 3.
That is why Eq. (\ref{eq:cusp}) does not contain
the cubic term.
We also note that, for the case of two-dimensional
control space, the structurally stable catastrophe
is classified as either the fold or cusp type by Thom's theorem.
Therefore, at point $a$, a second-order 
phase transition occurs.
If we fix $N$ to $N_a$,
\begin{equation}
  \omega= C^{-1/3}\varphi_0^{1/3}.
\end{equation}
This is stable.
From Eqs. (\ref{eq:cube})-(\ref{eq:critical}), with $\varphi_0=0$ 
near point $a$, it is derived that
the scalar susceptibility $\chi_{\varphi}$ is given by
\begin{equation}
  \chi_{\varphi}=\cases{
    D^{-1}(N_a-N)^{-1} & for $N<N_a$, \cr
    \frac{1}{2}D^{-1}(N-N_a)^{-1} & for $N>N_a$. \cr}
\end{equation}
Near point $a$
in region (A) in Fig. 3, only one real root
of the cubic equation (\ref{eq:cube}) 
corresponds to the stable equilibrium solution,
while, in region (B), the smallest and largest roots of
three real roots correspond to
the stable equilibrium solutions and the intermediate root 
corresponds to the unstable one.
If $\varphi_0>0$, the largest 
root corresponds to the globally stable one.
If $\varphi_0<0$, the smallest 
root corresponds to the globally stable one.
If $\varphi_0=0$, the two stable equilibrium solutions
have identical ADM energies. 
The bifurcation set $B_m$, 
which is the envelope $b^{\prime}ab$, is
given by
\begin{eqnarray}
  \left(\frac{\partial m}{\partial \omega}
  \right)_{\varphi_0,N}&=&
  C\omega^3-D(N-N_a)\omega-\varphi_0=0, \\
  \left(\frac{\partial^2 m}{\partial \omega^2}
  \right)_{\varphi_0,N}&=&
  3C\omega^2-D(N-N_a)=0,
\end{eqnarray} 
i.e.,
\begin{equation}
  \varphi_0=\pm\left(\frac{4D^3}{27C}\right)^{1/2}(N-N_a)^{3/2},
\end{equation}
near point $a$.
The cusp catastrophe has been named after this shape.
On the envelope $dcbab^{\prime}c^{\prime}d$
except for points $a$ and $d$, 
one of the two distinct stable equilibrium solutions,
the locally but not globally stable one,
disappears, and hence a first-order phase transition occurs
if the system obeys a perfect delay convention.
On envelope $ede^{\prime}$,
the stable equilibrium solution disappears.
The catastrophic feature on the envelopes 
except for point $a$
is the fold catastrophe described in the last subsection.
Point $d$ is not a cusp point but the intersection of two folds.

We should comment that, for the near critical case
$-4.9\alt\beta\alt -4.35$, the behavior of the
stable equilibrium solutions around the point of 
the maximum baryon number
is somewhat complicated, although
the structure of the cusp catastrophe at cusp
point $a$ is not changed.
Figures 7 and 8 show the curves $(\tilde{\rho}_c,N)$
with $\varphi_0=0$,
for $\beta=-4.5$ and $-4.85$, respectively.
For $-4.8\alt\beta\alt-4.35$, the number of stable equilibrium solutions
changes as 1, 2 (degenerate in Fig. 7), 3, 1, 0
as the control parameter $N$ is increased
continuously, as is seen in Fig. 7.
For $-4.9\alt\beta\alt-4.8$, the number of stable equilibrium solutions
changes as 1, 2 (degenerate in Fig. 8), 3, 2, 0 as $N$ is 
increased continuously,
as is seen in Fig. 8.
For $\beta\alt-4.6$ the maximum ADM energy with $\varphi_0=0$
is greater than that in general
relativity,
while, for $\beta\agt -4.6$, it is the same as that in general relativity.

The Kepler mass, which governs the Newtonian orbital motion of a test
body, is not the ADM energy $m$ in general, but~\cite{de1992}
\begin{equation}
  \tilde{\mu}=\frac{1+\alpha_0\alpha_A}{1+\alpha_0^2}m,
\end{equation}
where
\begin{equation}
  \alpha_A\equiv\frac{\partial\ln m}{\partial\varphi}=
  -\frac{\omega}{m}.
\end{equation}
When we consider the case of $\alpha_0=\beta\varphi_0=0$,
the Kepler mass is identical to the ADM energy. 
Therefore the argument above for $\varphi_0=0$ is also
valid for the Kepler mass.

Here we present the physical interpretation as to why spontaneous
scalarization occurs.
In spite of the absence of the potential in the Lagrangian,
the scalar field $\varphi$ obtains an effective potential term
$W(\varphi)$ which satisfies
\begin{equation}
  \frac{\partial W}{\partial \varphi}=-4\pi \alpha(\varphi) T_*
\end{equation}
because of the coupling with matter.
Note that $T_*$ depends on $\varphi$.
Then, if we consider $A(\varphi)$ of the form (\ref{eq:quadratic}),
\begin{equation}
  \frac{\partial V}{\partial\varphi}=-4\pi \beta\varphi T_*,
\end{equation}
and if $T_*= A^4(-\tilde{\rho}+3\tilde{p})$ is negative,
$\varphi=0$ is an unstable stationary point of the effective 
potential, if $\beta<0$.
On the other hand, the term from the spatial derivative
in Eq. (\ref{eq:fieldeq2})
has a contribution to stabilize the solution.
By these two competing effects, the stability of the
trivial configuration $\varphi=0$ against
spontaneous scalarization is governed.
For a detailed analysis
of the stability of the trivial configuration,
see~\cite{harada1997}.

If spontaneous scalarization occurs, the effective gravitational 
constant, which is $A^2(\varphi)=\exp(\beta\varphi^2)$ 
in the sense of the inverse
of the Brans-Dicke scalar field,
becomes considerably smaller than unity.
Thereby the gravitation becomes weaker
and a considerably larger mass than in general relativity
can be supported by the lower
matter pressure than in general relativity.

\section{Summary and Discussions}

The behavior of the equilibrium solutions of neutron stars 
in scalar-tensor theories of gravitation
shows a catastrophic feature, which is characterized by
a discontinuous change of the system.
When we consider a function $A(\varphi)$ of the form
$A(\varphi)=\exp(\frac{1}{2}\beta \varphi^2)$, the catastrophe
types are classified as fold and cusp catastrophes.
The appearance of the cusp catastrophe 
depends on whether $\beta\agt-4.35$ or $\beta\alt-4.35$.
From the fact that those types of catastrophes 
are structurally stable,
it is expected that they would be seen in a wide class of
coupling functions.

For $\beta\agt-4.35$, the fold catastrophe 
on the two-dimensional control space does occur.
The critical baryon number
and critical ADM energy
depend on $\varphi_0$.
For a baryon number smaller than the critical one,
one stable equilibrium solution exists, while, for a baryon number
larger than the critical one, no stable equilibrium solution exists.
In particular, for $\varphi_0=0$,
the stable equilibrium solution is completely identical to that in
general relativity.
The behavior of the scalar charge and 
scalar susceptibility near the critical 
baryon number is explained by the form of the potential function
of the fold catastrophe.

For $\beta\alt-4.35$, the cusp catastrophe
does occur while the fold catastrophe also occurs.
For $\beta\alt-4.9$,
there is some critical value of the scalar field, $\varphi_0^{crit}>0$.
If $|\varphi_0|>\varphi_0^{crit}$,
there is only one critical number $N^{crit1}$ that
depends on $\varphi_0$.
For $N<N^{crit1}$, one stable equilibrium solution exists,
while, for $N>N^{crit1}$, no stable equilibrium solution exists.
If $0<|\varphi_0|<\varphi_0^{crit}$, 
there are three critical baryon numbers
$N^{crit1}>N^{crit2}>N^{crit3}$.
For $N<N^{crit3}$ or $N^{crit2}<N<N^{crit1}$,
only one stable equilibrium solution exists.
For $N^{crit3}<N<N^{crit2}$, 
two distinct stable equilibrium solutions exist and
they do not agree with those in general relativity
even for the limit $\varphi_0\to 0$.
The almost general relativistic branch is unstable
for $N>N^{crit2}$.
For $N>N^{crit1}$, however, no stable equilibrium solution exists.
If $\varphi_0=0$, the sequence of equilibrium solutions of 
neutron stars bifurcates at a point.
Beyond this point, the general relativistic 
branch becomes unstable
and another two (degenerate) sequences of equilibrium solutions far from
the general relativistic one are stable.
This bifurcation point is a cusp point,
and the behavior of the scalar charge and
scalar susceptibility near the cusp point
is explained by the form of the potential
function of the cusp catastrophe.
At a point on the envelopes other than
the cusp point, the fold catastrophe occurs.
Since the critical baryon numbers $N^{crit1}$
and $N^{crit2}$ agree, the number of stable equilibrium solutions
is 1 for $N<N^{crit3}$, 2 for $N^{crit3}<N<N^{crit2}=N^{crit1}$
and 0 for $N^{crit2}=N^{crit1}<N$.
It should be noticed that, 
for the near critical case $-4.9\alt\beta\alt-4.35$, 
the structure of the cusp
catastrophe does appear although the behavior 
becomes somewhat more complicated around 
the maximum baryon number for $\varphi_0\simeq 0$.
This complicated feature agrees with the fact that 
the critical mass against zero-mode instability 
is not a monotonic function 
with respect to $\beta$,
which is seen in Table I of~\cite{harada1997}.

Here we comment on the continuous change 
of the asymptotic value of the scalar field $\varphi_0$. 
If we identify $\varphi_0$ with the cosmological value
of the scalar field,
the evolution of
$\varphi_0$ can be described by the equation of motion
(\ref{eq:fieldeq2})
in the Friedmann-Robertson-Walker universe.
On the other hand, 
if we identify $\varphi_0$ with the value of the scalar field
at the matching region in the $N$-compact-body problem,
$\varphi_0$ should evolve due to 
the change of the density distribution
around the neutron star.
If the time scale of the variation of $\varphi_0$ 
is sufficiently longer
than that of the local gravitational phenomena,
such as the scalar gravitational wave radiation,
the process due to the change of $\varphi_0$ can be 
regarded as quasistatic.
Through the cosmological evolution of the scalar field $\varphi_0$,
the neutron stars may collapse and radiate a scalar gravitational 
wave.

We also comment on the continuous change of $N$,
which may be a result of a mass accretion onto the neutron star.
If the baryon number of the neutron star exceeds the maximum 
value, the neutron star collapses and scalar gravitational 
waves are radiated and this is a candidate for the
source of the scalar gravitational 
waves~\cite{snn1994,sst1995,hcnn1997}.
In a theory like the one of Fig. 7, there is
a stable general relativistic neutron star that has the same baryon number and
ADM energy within numerical accuracy as the maximum-mass 
non-general-relativistic neutron star has.
Then, the transition of the non-general-relativistic neutron star 
to the general relativistic one due to a mass accretion
occurs without any energy extraction.

Scalar-tensor theories of gravity
naturally arise from the low-energy limit of 
string theory or other 
unified theories.
For the moment, however, 
it is not clear how the scalar fields
should couple to gravity (but see~\cite{dp1994}).
Experimental tests, such as binary
pulsar timing observations,
may constrain the way of coupling
between the scalar fields and gravity.
In particular, as for the case in which
the single, massless scalar field
couples to gravity with the coupling function
$A(\varphi)=\exp[(1/2)\beta\varphi^2]$,
Damour and Esposito-Far\`ese~\cite{de1996}
obtained the constraint on $\beta$ as
$\beta\agt -5$,
using the data of three binary pulsars.
They showed that the occurrence of 
spontaneous scalarization makes it very
difficult for the theory 
to maintain consistency with the results of
binary pulsar timing experiments.
The results obtained in this paper
show that spontaneous scalarization is
not an exceptional but robust phenomenon
for the neutron star and
common to a wide range of coupling functions.
Gravitational experiments with high-precision 
and/or in a strong-field regime
and gravitational wave observations
may have the potential 
to constrain the way of coupling of the gravitational
scalar fields and thereby we may catch a glimpse of 
string-scale physics.

\acknowledgements
I would like to thank T. Nakamura,
M. Sasaki,
Y. Eriguchi, N. Sugiyama, K. Nakao, 
M. Siino, T. Chiba,
and M. Kaneko for useful discussions.
I am also grateful to H. Sato
for his continuous encouragement.

\appendix

\newpage


\vskip 0.3in
\centerline{FIGURE CAPTION}
\vskip 0.05in

\newcounter{fignum}
\begin{list}{Fig.\arabic{fignum}.}{\usecounter{fignum}}

\item
A family of curves of $\tilde{\rho}_c=\mbox{const}$ in the 
$(\varphi_0,N)$ plane for the $\beta=-4$ case.
The ordinate is 
$m_b N$
in place of the baryon number $N$.
In region (A), only one stable equilibrium solution exists,
while, in region (B), no stable equilibrium solution exists.
At a point on the envelope of the family of the curves,
the fold catastrophe occurs.
\item
Curves $(\varphi_0,\omega)$ with $N$ constant 
for the $\beta=-4$ case.
The number attached to each curve is 
$m_{b}N$ in the solar mass unit.
The solid lines denote stable branches and
the dotted lines denote unstable branches.
\item
Same as Fig. 1, but for $\beta=-6$ case.
In region (A), only one stable equilibrium solution exists.
In region (B), two distinct stable equilibrium solutions exist.
In region (C), no stable solution exists.
Point $a$ is a cusp point.
At point $a$, the cusp catastrophe occurs,
while
the fold catastrophe occurs
at a point on the envelopes except for $a$.
\item
Same as Fig. 2, but for the $\beta=-6$ case.
\item
(a) $(\tilde{\rho}_c,m)$ and (b) $(\tilde{\rho}_c,N)$ curves
with $\varphi_0=0$ for $\beta=-6$.
The solid lines denote stable branches, while the dotted lines
denote unstable branches.
The two distinct bifurcated branches are degenerate
because they have identical ADM energies and 
baryon numbers but scalar fields of the opposite sign.
The number of stable equilibrium solutions changes as 1, 2, 0 as $N$ increases.\item
Equilibrium space $M_m$ in the
$(\varphi_0,N,\omega)$ space 
around cusp point $a$
for the $\beta=-6$ case.
This structure of the equilibrium space 
is classified as the cusp catastrophe.
\item
Same as Fig. 5(b), but for $\beta=-4.5$.
The number of stable equilibrium solutions changes as 1, 2, 3, 1, 0
as $N$ increases.
Two solutions are degenerate on the 
non-general-relativistic
branches.
\item
Same as Fig. 5(b), but for $\beta=-4.85$.
The number of stable equilibrium solutions changes as 1, 2, 3, 2, 0
as $N$ increases.
Two solutions are degenerate on the 
non-general-relativistic
branches.
\end{list}

\newpage
\begin{figure}
      \vspace{1cm}
      \centerline{\epsfysize 9cm \epsfxsize 13cm \epsfbox{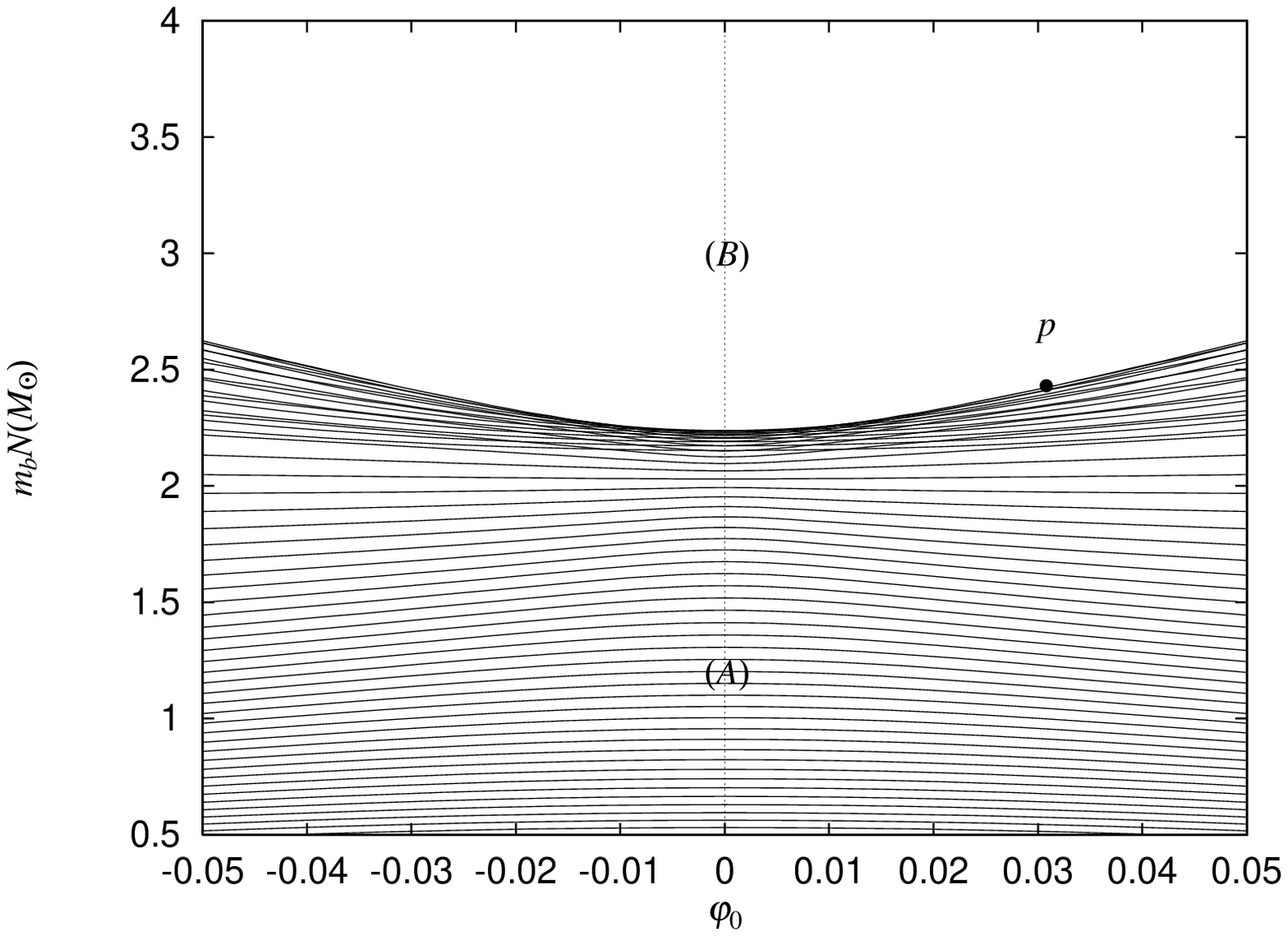}}
\caption{}
      \vspace{1cm}
      \centerline{\epsfysize 9cm \epsfxsize 13cm \epsfbox{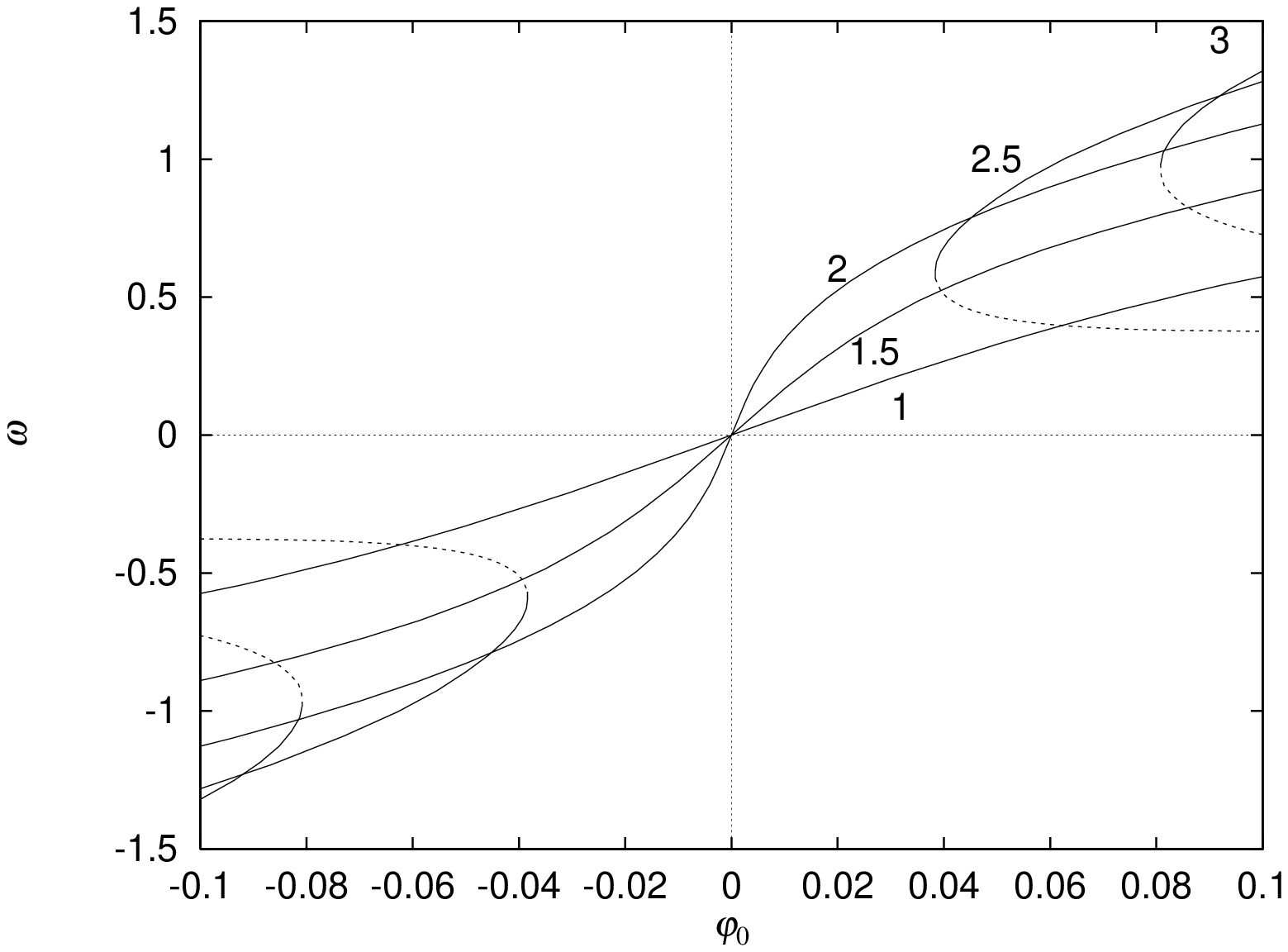}}
\caption{}
\newpage
      \vspace{1cm}
      \centerline{\epsfysize 9cm \epsfxsize 13cm \epsfbox{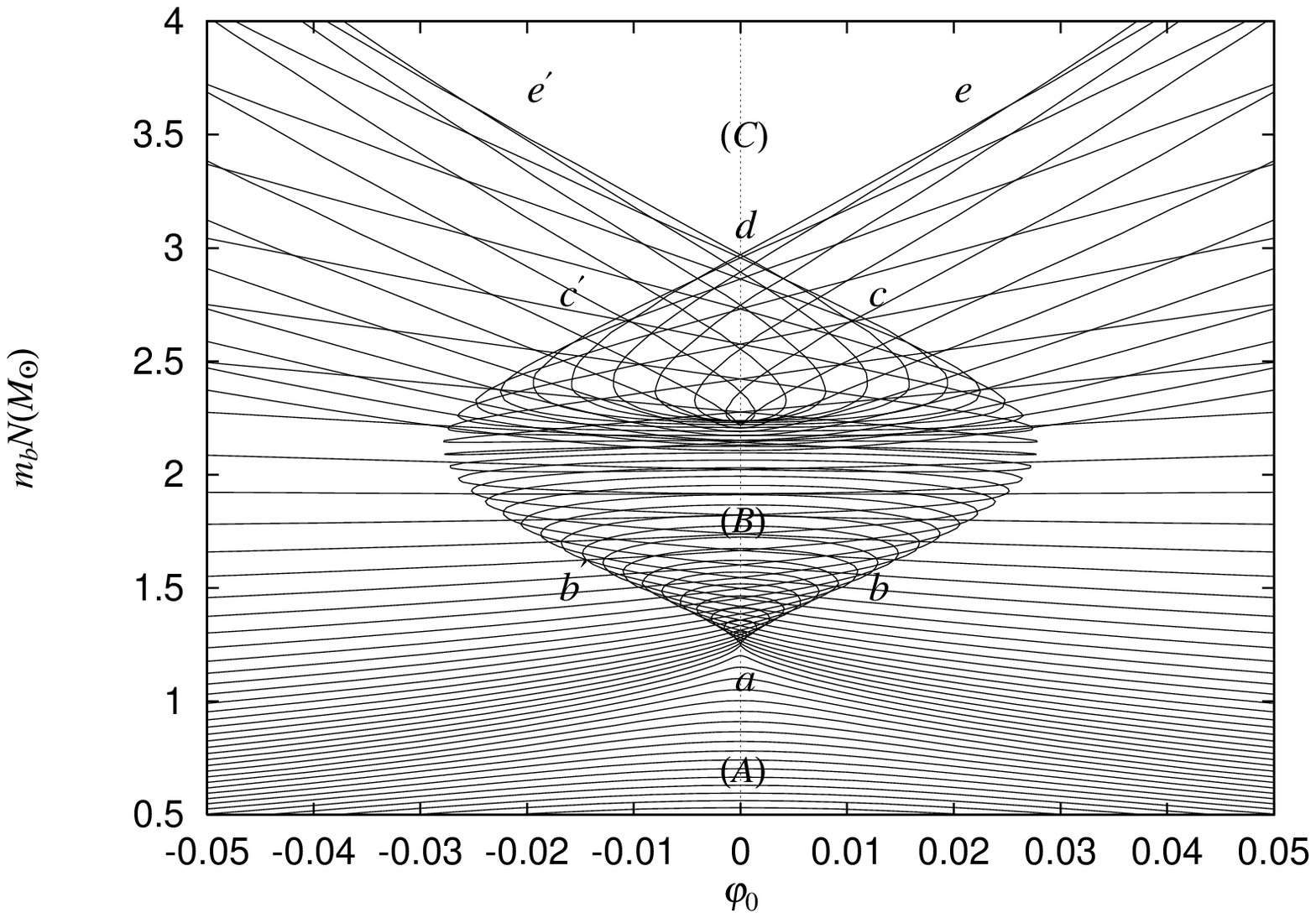}}
\caption{}
      \vspace{1cm}
      \centerline{\epsfysize 9cm \epsfxsize 13cm \epsfbox{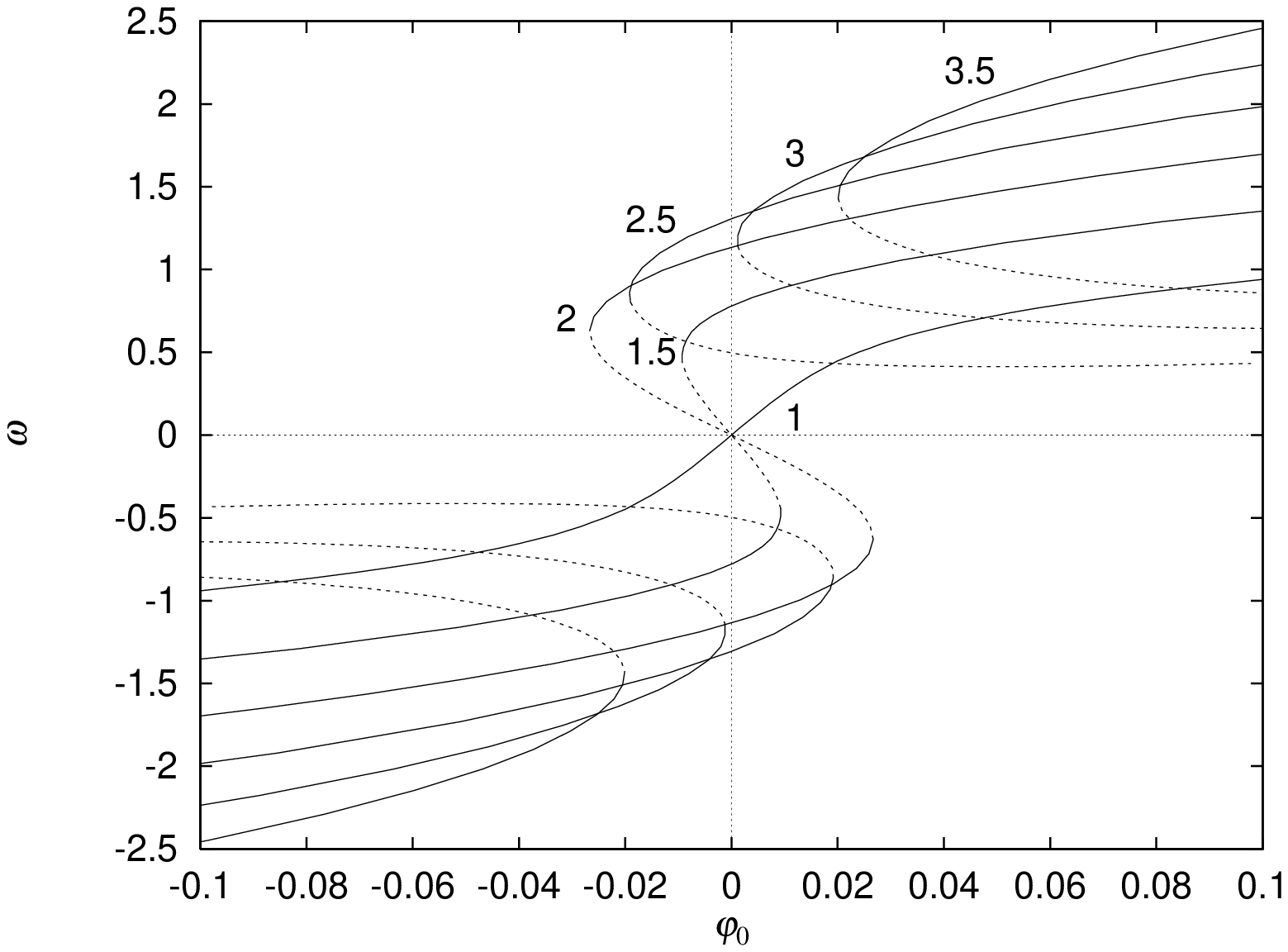}}
\caption{}
\newpage
      \vspace{1cm}
      \centerline{\epsfysize 9cm \epsfxsize 13cm \epsfbox{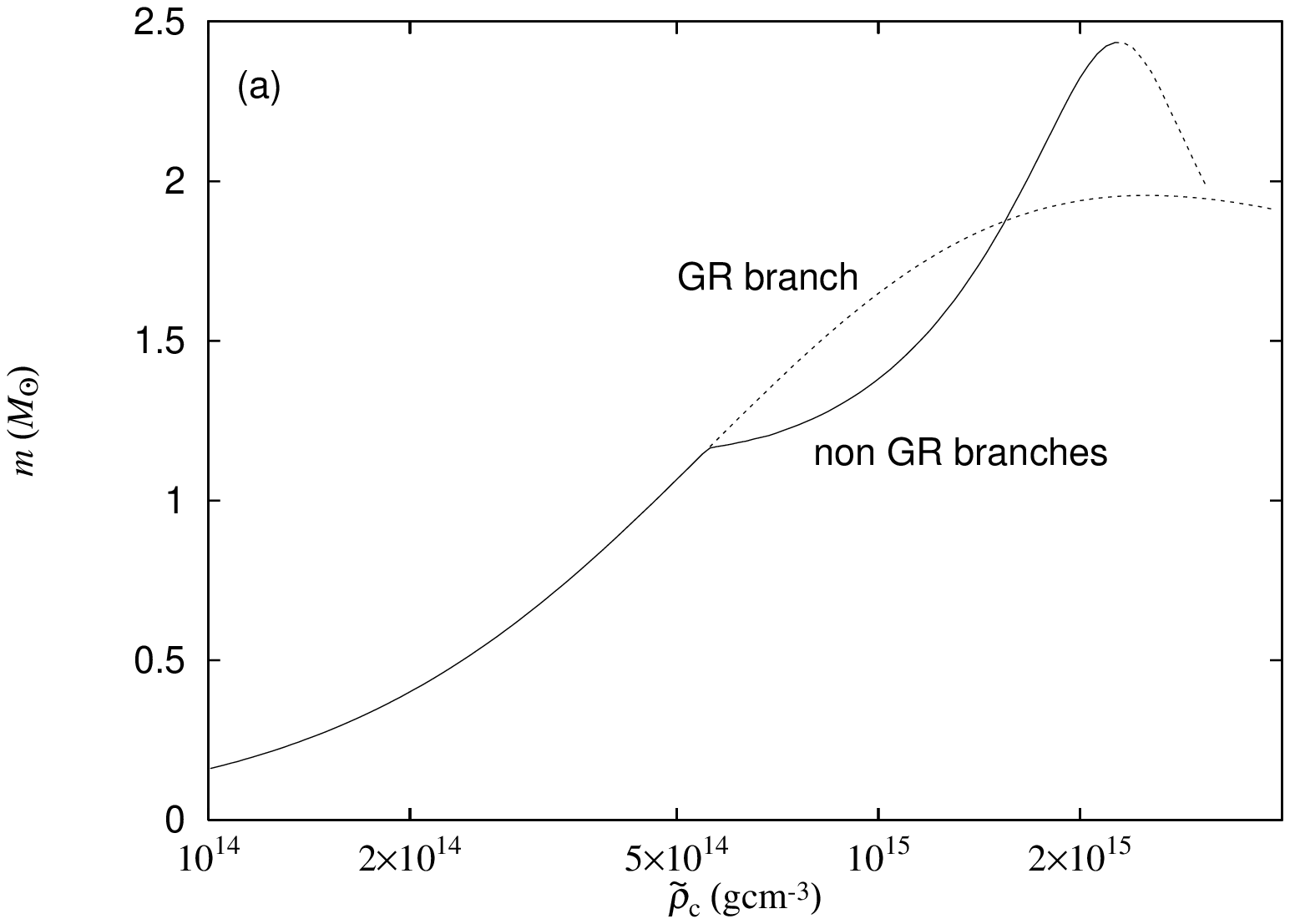}}
      \vspace{1cm}
      \centerline{\epsfysize 9cm \epsfxsize 13cm \epsfbox{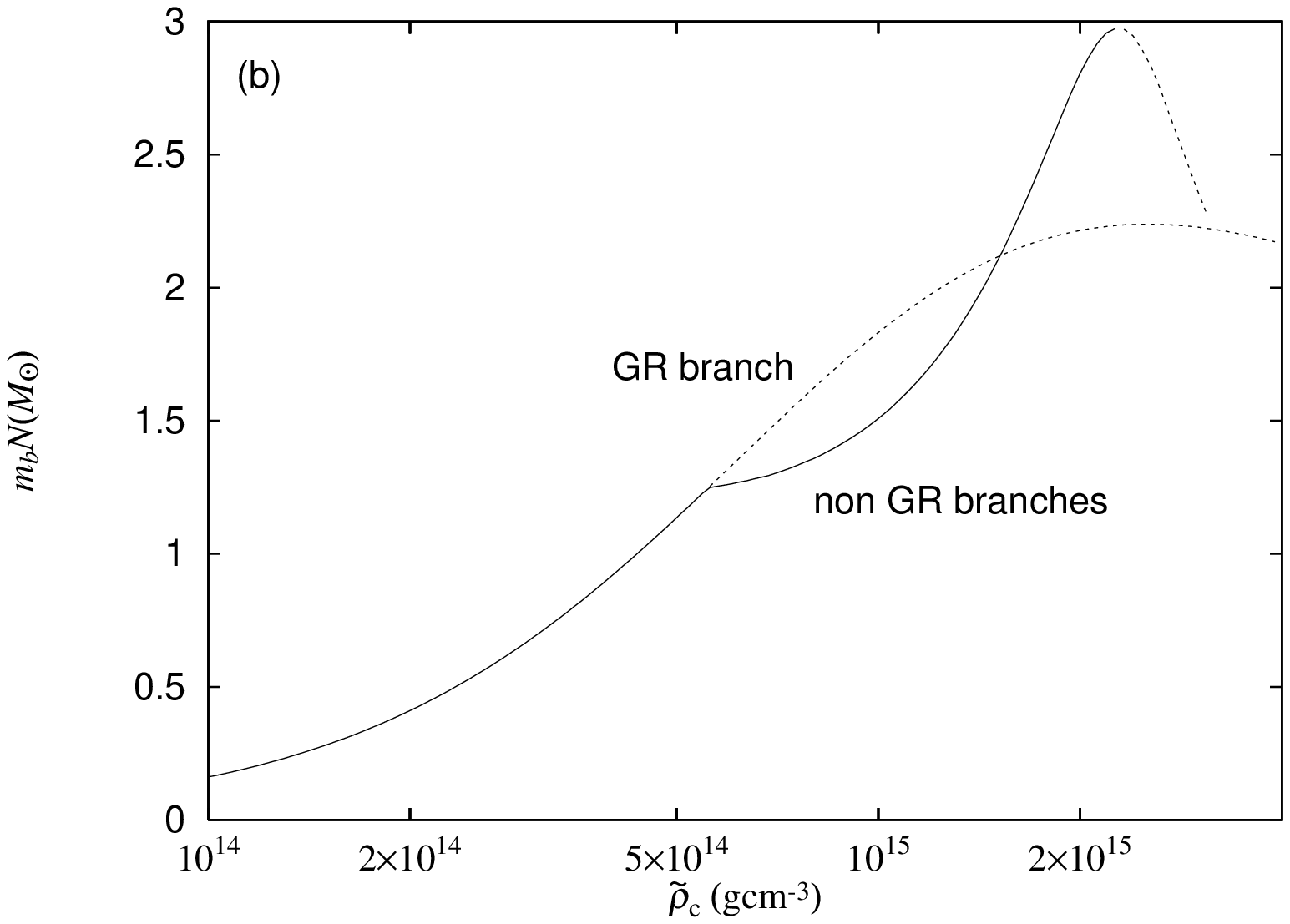}}
\caption{}
\newpage
      \vspace{1cm}
      \centerline{\epsfysize 9cm \epsfxsize 13cm \epsfbox{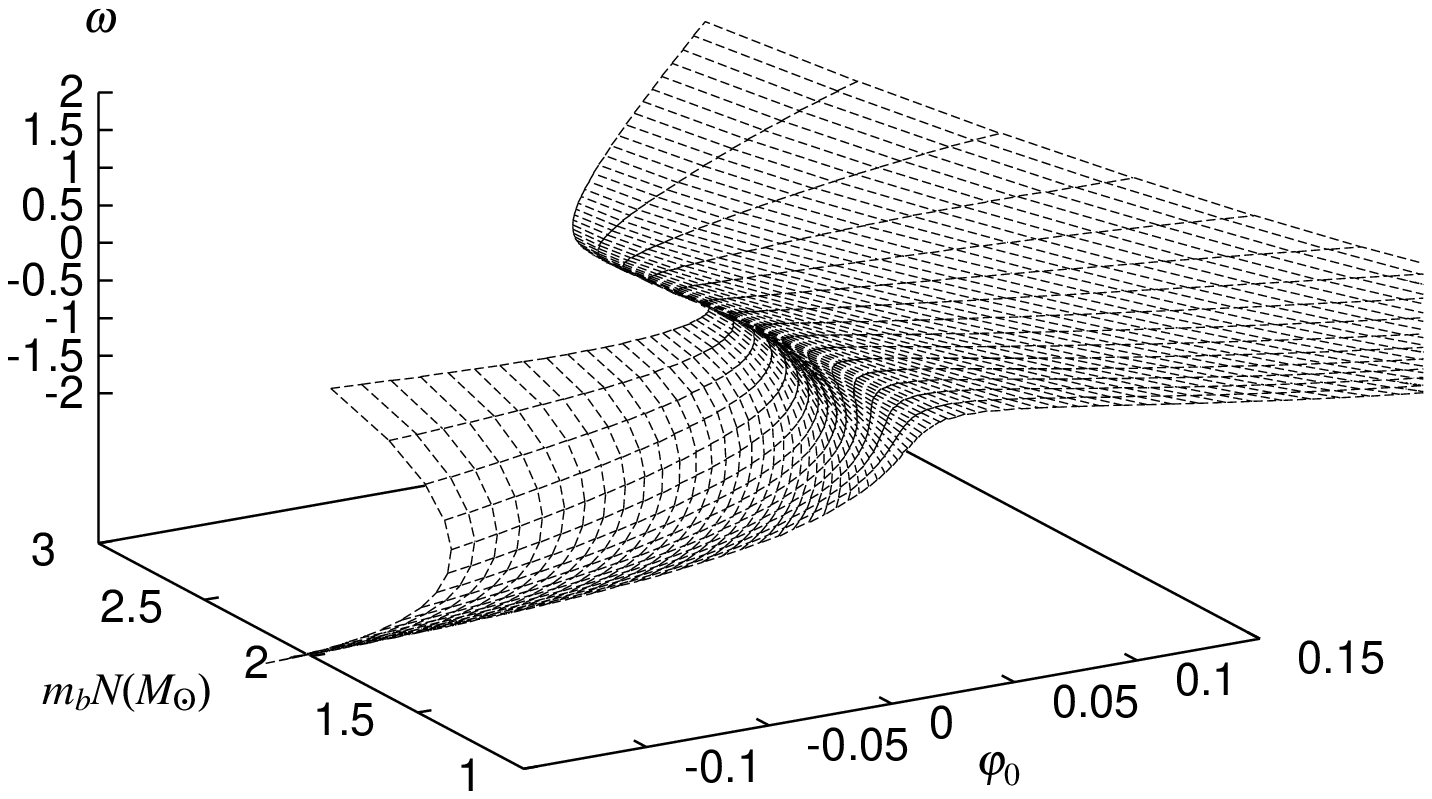}}
\caption{}
      \vspace{1cm}
      \centerline{\epsfysize 9cm \epsfxsize 13cm \epsfbox{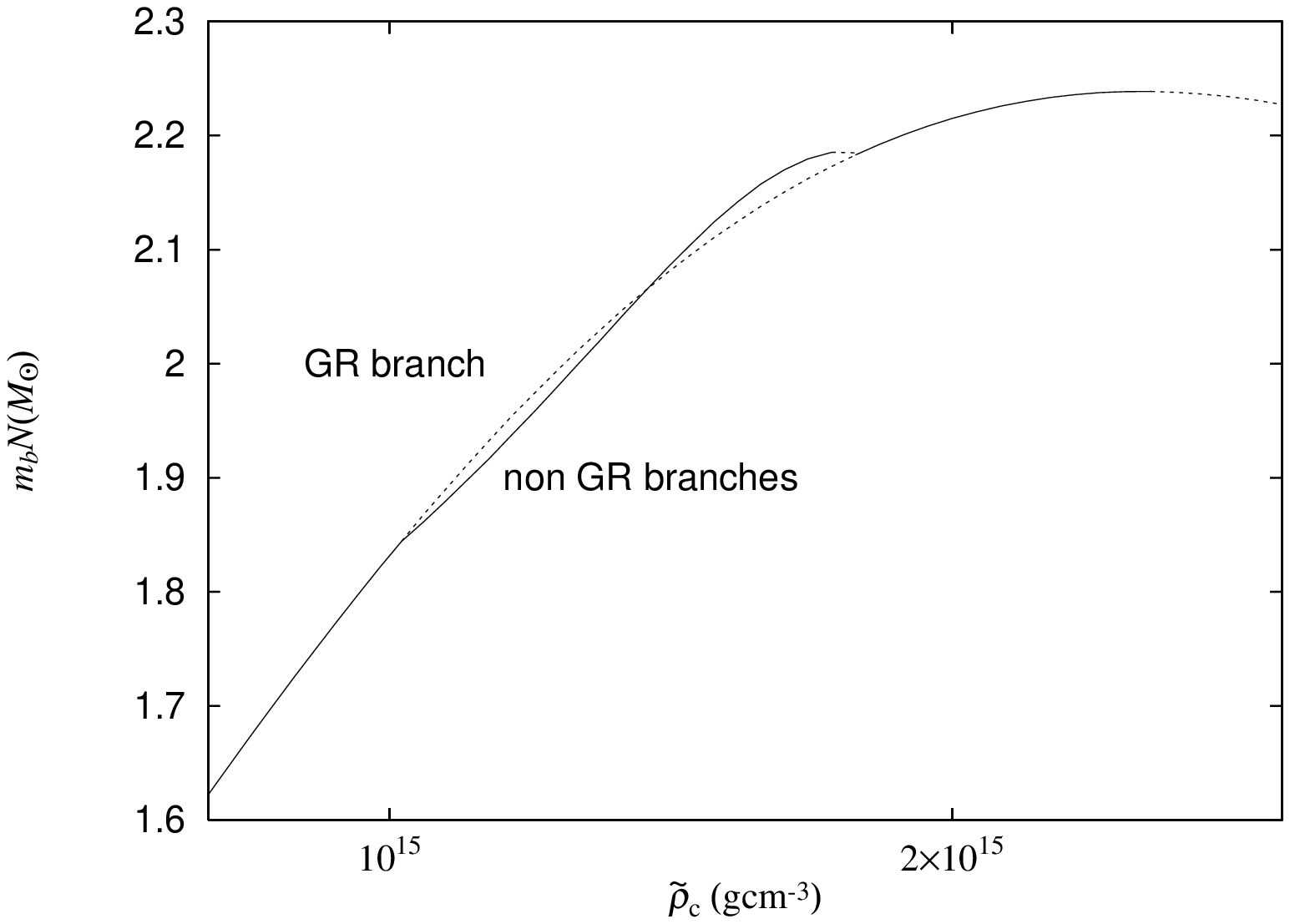}}
\caption{}
\newpage
      \vspace{1cm}
      \centerline{\epsfysize 9cm \epsfxsize 13cm \epsfbox{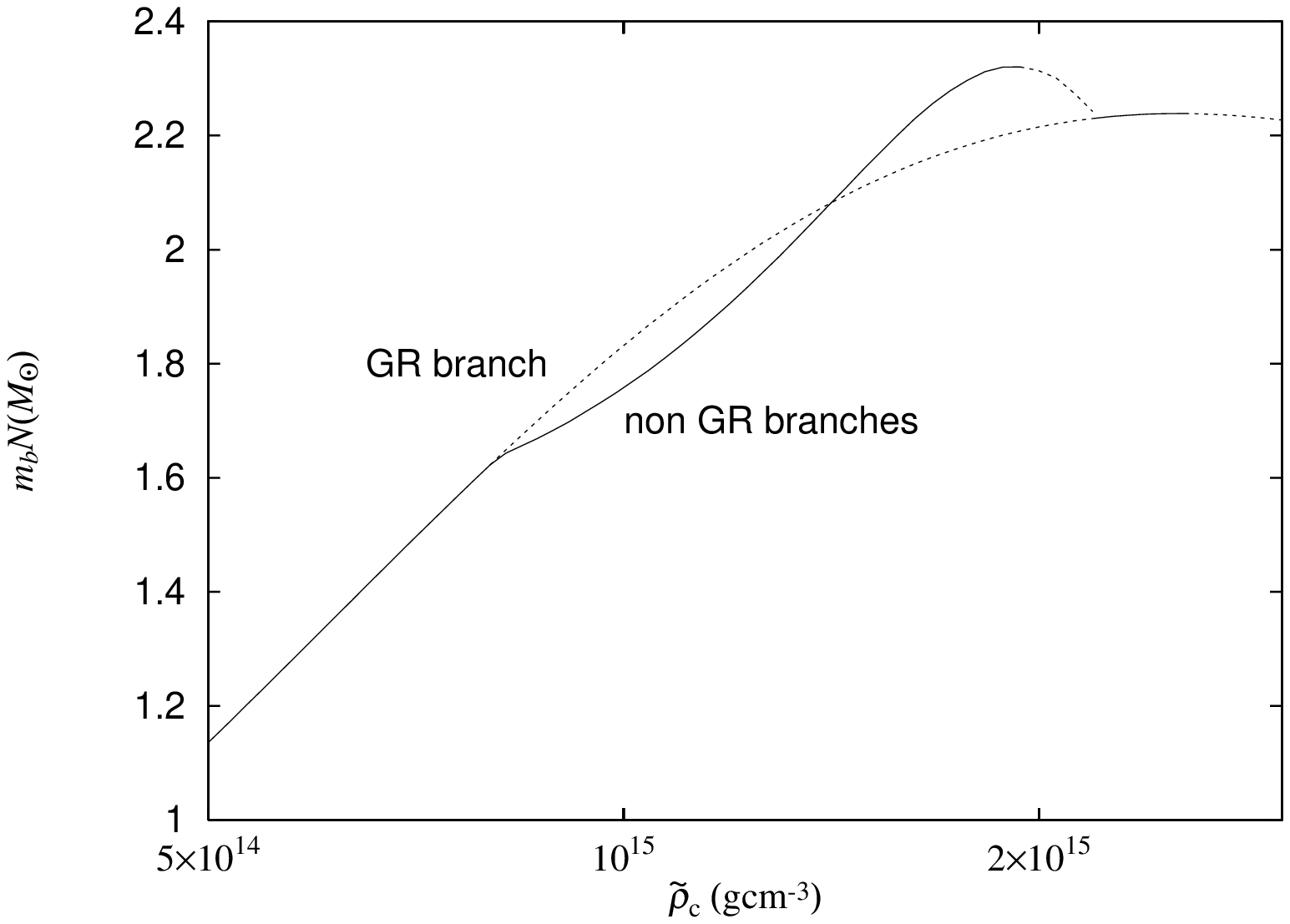}}
\caption{}
\end{figure}%

\begin{thebibliography}{99}
\bibitem{will1993}
  C. M. Will,
  {\it Theory and Experiment in Gravitational Physics},
  revised ed. (Cambridge University Press, Cambridge,
  England, 1993).
\bibitem{de1992}
  T. Damour and G. Esposito-Far\`ese, 
  Class. Quantum Grav.
  {\bf 9}, 2093 (1992).
\bibitem{bd1961}
  C.~Brans and R.H.~Dicke,
  Phys. Rev. 
  {\bf 124}, 925 (1961).
\bibitem{gsw1987}
  M.B.~Green, J.H.~Schwarz, and E.~Witten,
  {\it Superstring Theory},
  (Cambridge University Press, Cambridge, England, 1987),
  Vols. 1 and 2.
\bibitem{sa1990}
  P.J.~Steinhardt and F.S.~Accetta,
  Phys. Rev. Lett.
  {\bf 64}, 2740 (1990).
\bibitem{ligo}
  A.~Abramovici {\it et al.},
  Science,
  {\bf 256}, 325 (1992).
\bibitem{virgo}
  C.~Bradaschia {\it et al.}.
  Nucl. Instrum. Methods Phys. Res. A
  {\bf 289}, 518 (1990).
\bibitem{geo}
  J. Hough, 
    in
    {\it Proceedings of the Sixth Marcel Grossmann Meeting on General
    Relativity}, Kyoto, Japan, 1991,
    edited by H. Sato and T. Nakamura
    (World Scientific, Singapore, 1992), p. 192.
\bibitem{tama}
    K. Kuroda {\it et al.,} in
    {\it Proceedings of International Conference on
    Gravitational Waves: Sources and Detectors}, Pisa, Italy,
    , 1996, edited by I. Ciufolini and F. Fidecaro
    (World Scientific, Singapore, 1997), p. 100.
\bibitem{wz1989}
  C.M. Will, and H.W. Zaglauer,
  Astrophys. J. 
  {\bf 346}, 366 (1989).
   \bibitem{will1994}
    C.M. Will, 
    Phys. Rev. D
    {\bf 50}, 6058 (1994).
  \bibitem{snn1994}
    M. Shibata, K. Nakao, and T. Nakamura, 
    Phys. Rev. D
    {\bf 50}, 7304 (1994).
\bibitem{sst1995}
  M.A. Scheel, S.L. Shapiro, and S.A. Teukolsky,
  Phys. Rev. D
  {\bf 51}, 4208 (1995);
  {\bf 51}, 4236 (1995).
\bibitem{hcnn1997}
  T. Harada, T. Chiba, K. Nakao, and T. Nakamura,
  Phys. Rev. D
  {\bf 55}, 2024 (1997).     
\bibitem{de1993}
  T. Damour and G. Esposito-Far\`ese, 
  Phys. Rev. Lett.
  {\bf 70}, 2220 (1993).
\bibitem{de1996}
  T. Damour and G. Esposito-Far\`ese, 
  Phys. Rev. D
  {\bf 54}, 1474 (1996).
\bibitem{katz1978}
  J. Katz,
  Mon. Not. R. Astron. Soc.,
  {\bf 183}, 765 (1978).
\bibitem{katz1979}
  J. Katz,
  Mon. Not. R. Astron. Soc.,
  {\bf 189}, 817 (1979).
\bibitem{sorkin1981}
  R. Sorkin,
  Astrophys. J.
  {\bf 249}, 254 (1981).
\bibitem{sorkin1982}
  R. Sorkin,
   Astrophys. J.
  {\bf 257}, 847 (1982).
\bibitem{cs1997}
  G.L. Comer and H. Shin-kai,
  gr-qc/9708071, Class. Quantum Grav. (to be published).
\bibitem{tls1997}
  D.F. Torres, A.R. Liddle, and F.E. Schunk,
  gr-qc/9710048.
\bibitem{harada1997}
  T. Harada,
  Prog. Theor. Phys.
  {\bf 98}, 359 (1997).
\bibitem{mtw1973}
  C.W.~Misner, K.S.~Thorne, and J.A.~Wheeler,
  {\it Gravitation}
  (Freeman, New York, 1973).
\bibitem{lebach1995}
      D.E.~Lebach {\it et al.},
      Phys.~Rev.~Lett.\
      {\bf 75}, 1439 (1995).
\bibitem{williams1995}
  J.G. Williams, X.X. Newhall, and J.O. Dickey,
  Phys. Rev. D
  {\bf 53}, 6730 (1996).
\bibitem{htww1964}
  B.K. Harrison, K.S. Thorne, M. Wakano, and
  J.A. Wheeler,
  {\it Gravitation Theory and Gravitational Collapse}
  (The University of Chicago Press, Chicago, 1964).
\bibitem{chandrasekhar1964}
  S. Chandrasekhar,
  Astrophys. J.
  {\bf 140}, 417 (1964).
\bibitem{dp1994}
  T. Damour and A.M. Polyakov,
  Nucl. Phys. 
  {\bf B423}, 532 (1994);
  Gen. Relativ. Gravit.
  {\bf 26}, 1171 (1994).
\end{thebibliography}
\end{document}